# Ferromagnetic dynamics detected via one- and two-magnon NV relaxometry

Brendan A. McCullian[1], Ahmed M. Thabt[1], Benjamin A. Gray[2], Alex L. Melendez[1], Michael S. Wolf[2], Vladimir L. Safonov[2], Denis V. Pelekhov[1], Vidya P. Bhallamudi[3], Michael R. Page[2], P. Chris Hammel[1]

[1] – *Department of Physics, Ohio State University*
[2] – *Materials and Manufacturing Directorate, Air Force Research Laboratory*
[3] – *Department of Physics, Indian Institute of Technology Madras*

## Abstract

The NV center in diamond has proven to be a powerful tool for locally characterizing the magnetic response of microwave excited ferromagnets. To date, this has been limited by the requirement that the FMR excitation frequency be less than the NV spin resonance frequency. Here we report NV relaxometry based on a two-magnon Raman-like process, enabling detection of FMR at frequencies higher than the NV frequency. For high microwave drive powers, we observe an unexpected field-shift of the NV response relative to a simultaneous microwave absorption signal from a low damping ferrite film. We show that the field-shifted NV response is due to a second order Suhl instability. The instability creates a large population of non-equilibrium magnons which relax the NV spin, even when the uniform mode FMR frequency exceeds that of the NV spin resonance frequency, hence ruling out the possibility that the NV is relaxed by a single NV-resonant magnon. We argue that at high frequencies the NV response is due to a two-magnon relaxation process in which the difference frequency of two magnons matches the NV frequency, and at low frequencies we evaluate the lineshape of the one-magnon NV relaxometry response using spinwave instability theory.

## Introduction

Nitrogen-vacancy (NV) centers in diamond are spin defects which are sensitive to magnetic fields and whose spin state can be detected optically [1-4]. NV centers have long spin coherence times at room temperature allowing for sensitive magnetometry and their atomic-scale size allows for high spatial resolution [2, 4-9]. Applied to ferromagnetic dynamics, NV magnetometry has been used to detect driven ferromagnetic resonance (FMR) [10], and long range coherent NV-spinwave coupling [11, 12]. A complimentary technique called NV relaxometry, where changes to the NV spin lifetime as a result of magnetic field noise at the NV resonant frequency, has been employed as a sensitive tool for spectroscopic studies of ferromagnetic spinwaves [10, 13-19], electron spins in aqueous solution [20], nonequilibrium current fluctuations in graphene [21], and paramagnetic spins in diamond [22, 23]. Thus far, direct NV measurements of magnetic noise from ferromagnets have been limited to few-GHZ because of the requirement that the NV frequency be higher than the driven FMR frequency. The uniform mode FMR is the lowest frequency ferromagnetic excitation for a thin film, meaning that if the FMR frequency exceeds the NV frequency then there are no possible spinwaves available which are frequency matched to the NV sensor.

Recently, it was proposed that magnetic field noise from pairs of magnons in a ferromagnet could relax NV centers [24], provided that the ferromagnet can be driven to a large magnon chemical potential, which means an excess population of magnons above the thermal occupation level. Microwave drive of a ferromagnet on resonance has been shown to raise the magnon chemical potential [13]. This experimental result and the theoretical prediction of two-magnon NV relaxation provide a pathway to drive and ultimately detect ferromagnetic spin dynamics at above-NV frequencies.

Here we report driven FMR spectroscopy of a thin film of low-damping nickel zinc aluminum ferrite ($Ni_{0.65}Zn_{0.35}Al_{0.8}Fe_{1.2}O_4$, NZAFO). We simultaneously detect conventional broadband microwave absorption and NV relaxometry, finding that the two techniques have different spectral response. We explore two regimes of NV response: when then ferromagnetic uniform mode frequency is below the NV spin resonance frequency, and the converse. Using conventional microwave absorption, we demonstrate that NZAFO undergoes a second-order spinwave instability at high microwave drive powers. This driven spinwave instability is responsible for populating the magnons well above the thermal level, and explains the spectral response of the NV in the first regime. In the second regime, we find that the spinwave instability produces sufficient magnon population to relax the NV even when the NV spin resonance frequency is less the FMR frequency, a definitive detection of two-magnon NV relaxometry.

## Results

**Broadband NV Relaxometry**

NV relaxometry measures local magnetic field noise by optically detecting the spin state of fluorescent NV centers. NV centers are spin-1 defects in diamond which can be polarized into a maximally fluorescent spin-0 state under green illumination. The fluorescence intensity of an NV center or an ensemble of NVs is determined by a competition between the laser polarization rate and the longitudinal spin lifetime. If an NV is brought near a ferromagnet, dipole magnetic field noise from thermally occupied, spatially inhomogeneous ferromagnetic spinwaves can couple to the NV ground state spin transitions and shorten the effective NV spin lifetime, provided that the spinwaves are frequency matched to the NV. This effective spin lifetime now sets the NV fluorescence intensity for the NV-on-ferromagnet system. In our NV relaxometry experiment we compare the fluorescence intensity of NVs deposited on a ferromagnetic film when the ferromagnet is resonantly excited with microwaves with the fluorescence intensity when the ferromagnet is undriven.

In most previous NV relaxometry experiments [13, 15, 17-19] the uniform mode $(\vec{k} = 0)$ ferromagnetic resonance (FMR) was driven. The driven uniform mode FMR scatters off thermally occupied spinwaves $(\vec{k} \neq 0)$ in a four-magnon scattering, leading to an elevated population of various frequency and wavevector spinwaves [13]. This four-magnon scattering is responsible for converting spinwave population at the microwave driven uniform mode FMR frequency and producing NV-resonant spinwaves. Thus, it is useful in NV relaxometry to study ferromagnets with low damping, since for low damping ferromagnets the driven ferromagnetic mode will undergo significantly more magnon-magnon scattering before eventually damping to vibrational modes.

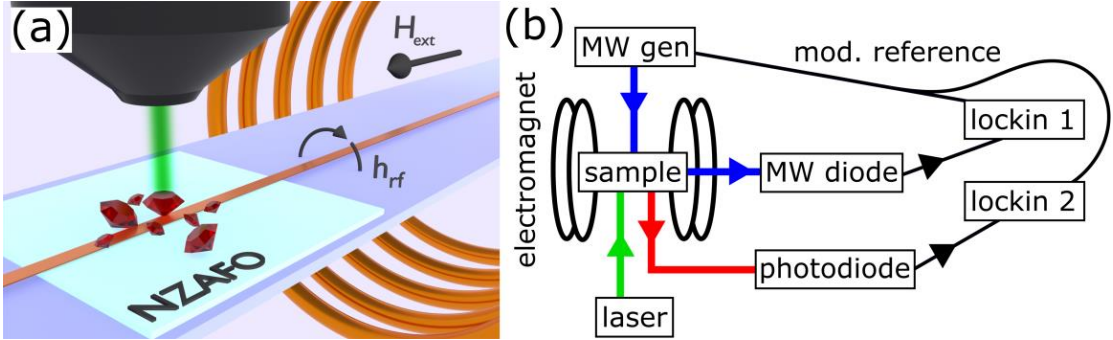

*Figure 1: NV relaxometry and microwave absorption experiment schematics*

*(a) Experimental schematic. In-plane magnetic field is swept while microwaves from microstrip drive NZAFO magnetization. Microscope objective focuses green polarizing laser onto NV nanodiamond powder and collects red NV fluorescence. (b) Detection schematic. Microwaves are amplitude modulated, NV fluorescence and microwave transmission are recorded by lockin amplifiers referenced to this modulation.*

We perform a continuous wave NV relaxometry and microwave absorption detection as outlined in **Figure 1 (a,b)**. The green laser used to polarize NVs is continuously on and addresses a few-hundred nm thick layer of nanodiamonds, each containing a few hundred NV centers which is drop cast on the microstrip and NZAFO surface. A microstrip antenna patterned on the surface of the ferromagnetic NZAFO film (see Methods) is used to resonantly excite ferromagnetic dynamics in the NZAFO. Microwaves are amplitude modulated and a lockin amplifier is used to detect changes in the NV fluorescence intensity at the microwave modulation frequency. External magnetic field $H_{\text{ext}}$ is swept in-plane. The microwave transmission is simultaneously monitored, allowing for direct comparison of the NV relaxometry and microwave absorption spectra.

**NV Relaxometry of NZAFO**

NZAFO is an insulating spinel ferrite with very low damping ($<3 \times 10^{-3}$) and saturation magnetization $4\pi M_s$ of 1500 Gauss which can be deposited epitaxially in thin films using pulsed laser epitaxy [25]. Our NZAFO film is deposited to a thickness of 23 nm on MgAl$_2$O$_4$ substrate and has a strong easy-plane magnetic anisotropy field ($H_{\text{ME}}$ ~1 T) due to lattice strain and also a modest ($H_{\text{cub}}$ ~40 Gauss) four-fold in-plane magnetic anisotropy field where ⟨110⟩ is the in-plane easy axis. All measurements in this work are with the static magnetic field applied in-plane, along the hard in-plane direction ⟨100⟩. The NZAFO FMR resonance frequency $f_{\text{res}}$ is given by **Equation 1**.

$$f_{res} = \frac{\gamma}{2\pi}\sqrt{(H_{res} - H_{cub})(H_{res} - H_{cub} + 4\pi M_s + H_{ME})} \tag{1}$$

**Figure 2** shows the NV detected ODFMR and microwave absorption spectra for three different input microwave powers. Microwave powers listed are the nominal output power at the microwave generator at 4.5 GHz. For all other frequencies the microwave power was decreased to maintain constant microwave transmission amplitude at each frequency (see Supplementary Material). NV signals are reported as percent which is calculated by taking the lockin detected NV fluorescence change and dividing by the DC photoluminescence at each field and frequency. We see both the NV ground state and excited state fluorescence response to direct microwave drive, which are not directly of interest in this work. The important NV signal is the fluorescence response due to the NZAFO ferromagnetic resonance. Simultaneously detected microwave absorption of the NZAFO is overlaid in white circles and disperses in accordance with **Equation 1**. At low microwave power the NV signal and the microwave absorption from NZAFO are spectrally very similar. With increasing microwave amplitude, we see that the NV response to the driven NZAFO shifts to lower magnetic field as compared with the microwave absorption peak. As magnetic field is increased the NZAFO FMR frequency crosses the NV ground state spin resonance frequency at around 150 Gauss. The NV relaxometry signal persists even after this crossing when the microwave amplitude is very high, and has not been discussed in previous NV measurements of driven ferromagnetic dynamics.

In order to understand the NV relaxometry signal we divide our analysis into two regimes: first, when the uniform mode FMR frequency is less than the NV spin resonance frequency, and conversely, when the NV spin resonance frequency is less than the uniform mode FMR frequency. The first regime corresponds to one-magnon relaxometry and the second regime is the two-magnon relaxometry. First though, it is important to understand the microwave power dependent ferromagnetic dynamics in NZAFO using conventional microwave absorption.

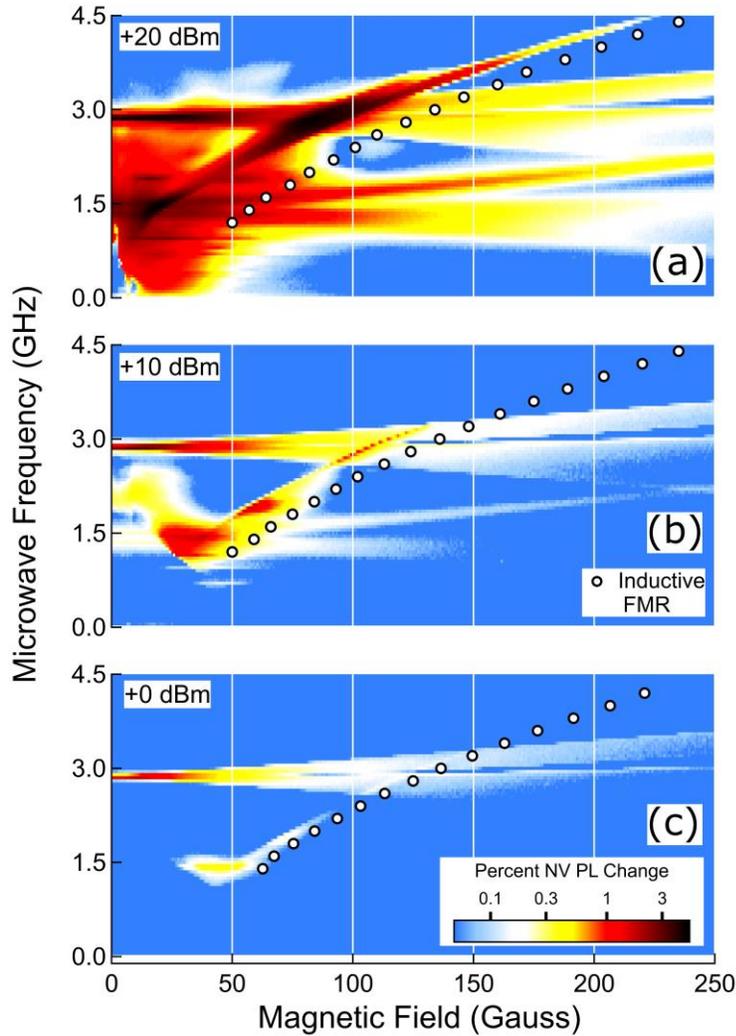

*Figure 2: Power dependent NV relaxation and microwave absorption*

*Field-swept NV relaxation spectra with the field swept along the NZAFO ⟨100⟩ axis for microwave input powers of +20 dBm (a), +10 dBm (b), and 0 dBm (c). Peak microwave absorption signal overlaid in white circles. NV relaxation plotted as percent change in photoluminescence (PL) by taking the NV lockin response and dividing by the DC NV PL at each point.*

**Second order spinwave instability in NZAFO**

In **Figure 3** we perform conventional broadband FMR of the NZAFO by measuring microwave absorption of the NZAFO film during magnetic field sweeps with a microwave excitation frequency of 4 GHz. We find that as microwave power is increased, the FMR line broadens and shifts toward lower magnetic field. We eliminate heating effects as contributions to this shift and broadening by performing microwave duty cycle and magnetic field sweep rate dependences (see Supplementary Material).

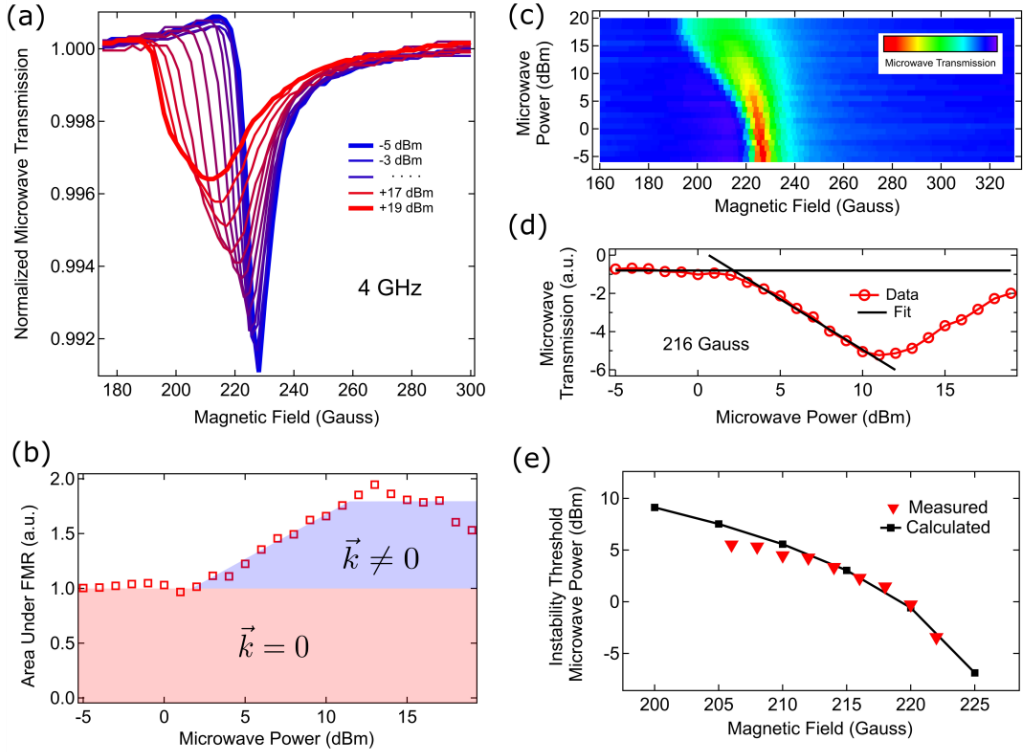

*Figure 3: Second-order spinwave Instability in NZAFO*

*(a) 4 GHz microwave absorption spectra with static magnetic field swept along the ⟨100⟩ NZAFO axis at varying microwave powers. Transmission normalized to unity off-resonance. (b) Area under normalized microwave absorption peaks in panel (a) vs microwave power. (c) Same absorption data as in panel (a), plotted explicitly vs microwave power to show the evolution of the FMR shoulder with increasing power. (d) Vertical linecut of panel (c) showing microwave transmission vs. microwave power at 216 Gauss. Intersection point of low-power constant microwave transmission with linear high-power transmission is the instability threshold for this field. This threshold is the lowest microwave power needed to drive the nonzero wavevector spinwave at this magnetic field and frequency in the instability process. (e) Spinwave instability threshold microwave power vs. magnetic field. Calculated instability threshold power overlaid.*

The broadening of the FMR line and the shift to lower magnetic field indicate the presence of a second order Suhl instability [26-30], where above a critical microwave power threshold, nonlinear coupling between the uniform FMR mode and the spinwaves results in an exponential increase in the population of finite-$\vec{k}$ magnons [31]. In the second order instability process two microwave photons whose frequency is close to the uniform mode FMR frequency are converted into pairs of spinwaves. We find in **Figure 3 (b)** that the area under the normalized FMR absorption curve increases with increasing power. Normalized here means that for each microwave power, the microwave transmission signal is normalized off resonance. In a microwave absorption experiment the area under the absorption curve should be proportional to the number of polarized spins. The NZAFO is fully polarized at even a few Gauss [25]. We therefore attribute the additional area under the microwave absorption curve to be from direct microwave drive of nonzero-$\vec{k}$ spinwaves.

In **Figure 3** we compare the measured spinwave instability threshold to calculation. This threshold is the lowest microwave power needed to drive the nonzero wavevector spinwave at this magnetic field and frequency in the instability process. The detected threshold at each magnetic field is determined by the microwave transmission as a function of power at that field. This can be understood as a vertical linecut in **Figure 3 (c)** which is the same data as **Figure 3 (a)** plotted to show the microwave transmission versus power more clearly. A representative cut of microwave transmission versus power for a magnetic field of 216 Gauss is shown explicitly in **Figure 3 (d)**. The instability threshold is determined by the intersection of the constant microwave transmission at low power with the linear decrease in transmitted power after the threshold. The calculated the instability threshold [31] (see Supplementary Material) is in good agreement with the threshold extracted from measurement in **Figure 3 (e)**.

## NV Relaxometry Regime 1: $\omega_{FMR} < \omega_{NV}$

In **Figure 4 (a)** we plot field-swept NV and microwave absorption signals at 2.2 GHz and +3 dBm. The microwave absorption signal shows a Lorentzian peak and a low-field shoulder consistent with the spinwave instability. The NV is relaxed at the same magnetic fields as instability shoulder, reaching a maximum intensity at the lowest magnetic field where the spinwave instability is driven.

The spectral response of the NV signal in the one-magnon relaxometry case is determined by the spinwave instability and the four-magnon scattering rate. The amplitude of the NV signal depends on four-magnon scattering which converts magnons at the microwave drive frequency into NV-resonant magnons. The four-magnon scattering rate goes approximately as $k^2$ [32]. Therefore, the NV signal should be maximal for the lowest field where the instability is driven, since this field corresponds to the largest $k$ that can be driven by the microwave field which ultimately results in the largest population of NV-resonant magnons.

In **Figure 4 (b)** we calculate dispersion of the spinwave manifold and the NV frequency range for magnetic fields corresponding to key features in the NV response in **Figure 4 (a)**. Beginning at high magnetic field (200 Gauss) we see that the microwave drive cannot couple to FMR or the spinwaves since there are no spinwaves at the microwave drive frequency. At 93 Gauss the uniform mode FMR is on resonance with the microwave drive. As magnetic field is decreased to 80, and then 65 Gauss the $|\vec{k}|$ of the instability-driven magnons is increased. This increase in the driven $|\vec{k}|$ results in more four-magnon scattering, which ultimately increases the number of NV resonant magnons. Eventually as magnetic field is decreased to 60 Gauss the microwave power becomes insufficient to overcome the spinwave instability threshold, and there are no driven magnons. This gives rise to the sharp cutoff of the NV signal between 65 and 60 Gauss. In spinwave theory the lowest microwave power threshold for driving the spinwave instability is for the $\vec{k} \parallel \vec{M}$ branch [32]. Using the NV data in **Figure 4 (a)** and the calculated spinwave manifold we find the cutoff $|\vec{k}|$ to be about $4.2 \times 10^{-5}$/m for the microwave conditions in **Figure 4 (a)**.

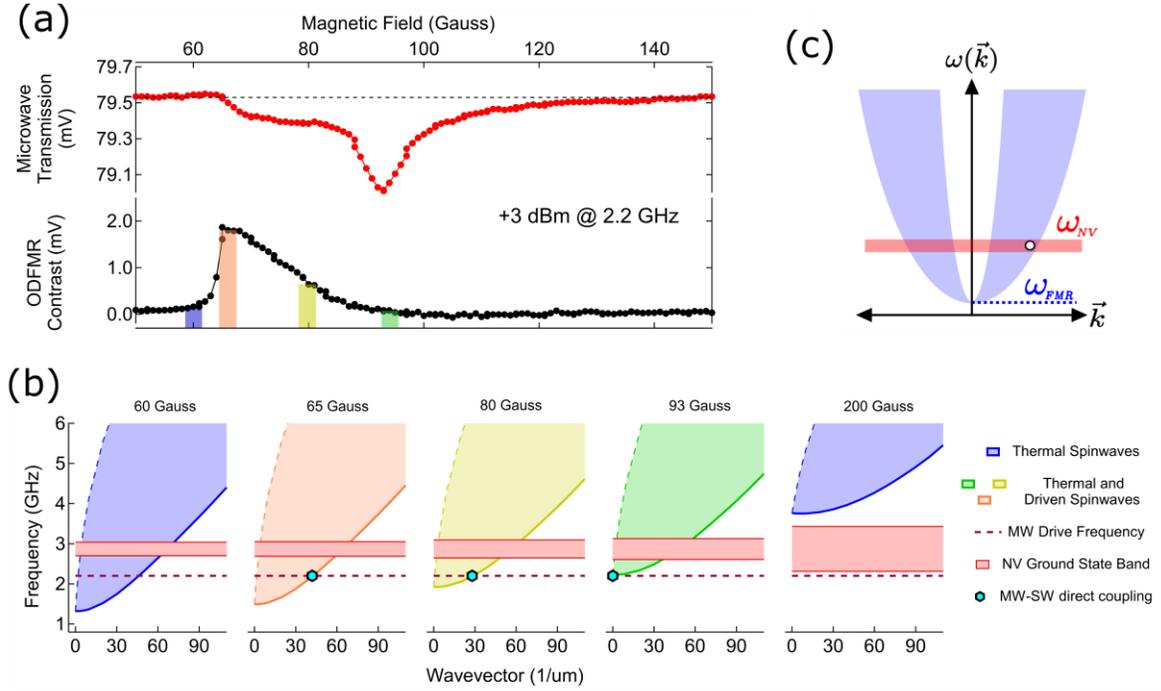

*Figure 4: One-magnon NV relaxometry*

*(a) Microwave absorption (red) and NV relaxation (black) signals vs. magnetic field at +3 dBm and 2.2 GHz. Four distinct magnetic fields highlighted in the NV response for comparison with calculation (colored bands). (b) Calculated spinwave states at 2.2 GHz for the magnetic fields of interest in panel (a), where spinwave envelope colors correspond to the colored bands in panel (a). Microwave drive frequency shown in red dotted line determines which spinwave (if any) is driven. Red band represents distribution of NV ground state frequencies for each applied field. Blue hexagons highlight the spinwave wavevector driven most efficiently by the microwave field, which are on the lower spinwave branch and have $\vec{k} \parallel \vec{M}$. (c) Schematic of one-magnon NV relaxometry. The ferromagnet is driven at low frequency near $\omega_{FMR}$ and magnon scattering populates the NV-resonant magnons (white circle) responsible for NV relaxation.*

### NV Relaxometry Regime 2: $\omega_{NV} < \omega_{FMR}$

In **Figure 5 (a)** we plot field-swept NV and microwave absorption signals at 4.0 GHz and +17 dBm. At this field and frequency, the uniform mode FMR frequency is above the NV frequency. As before, the microwave absorption signal shows a uniform mode FMR response and a low-field shoulder from the spinwave instability. The NV responds to the driven instability shoulder, despite the absence of any signal magnons which are frequency matched to the NV spin. The close spectral agreement of the NV response with the microwave absorption for the $\omega_{NV} < \omega_{FMR}$ case of **Figure 5 (a)** and **Figure 2 (a)** are proof that the ferromagnetic dynamics are responsible for this NV signal. This result is consistent with the proposed mechanism of two-magnon relaxometry [24] which suggests that a ferromagnet driven to large magnon chemical potential can host pairs of magnons which are frequency matched to the NV spin transition and cause relaxations, as shown schematically in **Figure 5 (b)**. In our experiment the spinwave instability is responsible for the highly elevated magnon population.

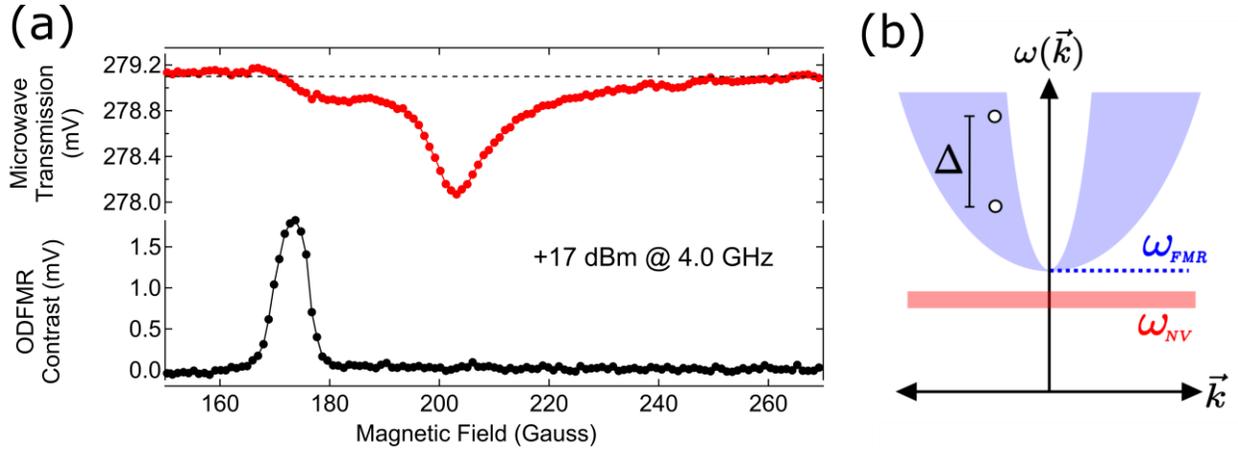

*Figure 5: Two-magnon NV relaxometry*

*(a) Microwave absorption (red) and NV relaxation (black) signals vs. magnetic field at +17 dBm and 4.0 GHz. NV response peaked strongly at the lowest magnetic field of the microwave absorption shoulder, which corresponds to the highest wavevector spinwave driven in the instability process. (b) Schematic of two-magnon NV relaxometry. The ferromagnet is driven near $\omega_{FMR}$ and magnon scattering populates pairs of magnons (white circles) responsible for NV relaxation. The difference frequency $\Delta$ of the two magnons must equal the NV frequency to cause relaxation.*

## Discussion

In summary, we have used NV relaxometry to detect driven ferromagnetic dynamics in NZAFO, demonstrating the first detection of two-magnon NV relaxometry. The second order spinwave instability in NZAFO allows us to increase the magnon chemical potential with increased microwave drive amplitude. Using an ensemble of NV nanodiamonds we optically detected the driven ferromagnetic spectroscopy, finding that the NV response has strong microwave power dependence. As the drive power is increased the NV signal is observed to disperse with the instability shoulder. For high drive power we find that the NV signal is present even when there are no single magnons frequency matched to the NV, demonstrating the presence of the two-magnon relaxometry process. We understand the lineshape of the one-magnon NV response in terms of the spinwave instability threshold and the four-magnon scattering rate which converts the instability-driven magnons into NV-resonant magnons.

There is reason to suspect that coupling pairs of magnons to NVs may work even at frequencies significantly higher than the NV frequency. The decreasing intensity of the NV relaxometry in our data has two contributions from the experimental parameters. First, to excite high frequency ferromagnetic dynamics the applied magnetic field is also increased, and this applied field will be off-axis for most of our randomly oriented NV ensemble. Off-axis magnetic field is known to degrade NV contrast because of NV-state mixing [33]. Second, the transmission of microwave stripline decreases with frequency, which limits our ability to drive the spinwave instability at high frequencies. If these experimental limitations can be overcome, then the ferromagnetic dynamics can be detected at frequencies potentially orders of magnitude larger than the NV spin frequency.

A next step would be to explore direct measurements of NV $T_1$ which may afford some additional sensitivity to small amounts of two-magnon noise. NZAFO and other insulating ferrite samples which are known to exhibit spinwave instabilities may play an important role in understanding the two-magnon relaxometry at very high frequencies.

We thank Phil Wigen for helpful discussion of spinwave instabilities. Funding for this research was provided primarily by the Center for Emergent Materials at The Ohio State University, a National Science Foundation (NSF) MRSEC through Award No. DMR-1420451, with partial support provided by the Army Research Office through Award No. W911NF-16-1-0547 and by the Air Force Office of Scientific Research (AFOSR) through Award No. FA9550-15RXCOR198. We acknowledge the use of Ohio State Nanosystems Laboratory shared facilities for device fabrication.

## Methods

### Materials

NZAFO was synthesized to a thickness of 23 nm onto single-crystal $MgAl_2O_4$ using pulsed laser epitaxy following the procedure outlined in reference [25]. NV nanodiamonds 10-200 nm in size were irradiated with a 1.5 MeV, $3.48 \times 10^{18}$/cm$^2$/h electron beam for 3 hours, creating a vacancy density of ~59 ppm. After subsequent annealing in a 4% $H_2$ in Ar environment and cleaning following the procedure following reference [34], nanodiamonds of ~30 ppm NV center concentration are obtained. Nanodiamonds were drop-cast on the NZAFO and microstrip to create a layer a few hundred nanometers thick.

### Broadband Microwave Absorption Spectroscopy

A tapered microstrip antenna with a taper region 150 µm long and 15 µm wide with composition Ti (5 nm) / Ag (285 nm) / Au (10 nm) was electron beam evaporated onto the surface of the NZAFO film. 100% amplitude modulated microwaves at a modulation frequency of 1.3 kHz were sent through microstrip in transmission and detected by a microwave diode. The microwave diode voltage was read by a lockin amplifier referenced to the modulation. Magnetic field was swept at constant microwave frequency.

### NV relaxometry

Fluorescence was excited in the NV centers using a 532 nm laser at a power of 30 mW, focused to a few micron wide spot using a 20X objective. The fluorescence detection path filtered out green laser with a dichroic mirror (Semrock 552 nm edge LaserMUX) and a long pass filter (Thorlabs FEL0600). Fluorescence was measured by a photodiode (Thorlabs DET110) amplified (RHK IVP300) and detected by a lockin amplifier referenced to the microwave amplitude modulation. The NV fluorescence signal was collected from nanodiamonds on top of the microstrip, and was qualitatively similar to data collected from nanodiamonds on the NZAFO and to the side of the microstrip.

**Calculating the Spinwave Dynamics**

To calculate the spinwave dynamics we follow the method outlined in reference [31]. The dynamics of the magnetization $\vec{M}$ are determined by the torque on $\vec{M}$ by the total effective magnetic field $\vec{H}_{\text{eff}}$

$$\frac{d\vec{M}}{dt} = -|\gamma|\vec{M} \times \vec{H}_{\text{eff}}$$

We expand the magnetization and field terms into their Fourier components, and include relevant magnetic anisotropies, dipolar effects, exchange effects, and the microwave drive field. From the torque equation we are able to calculate both the spinwave instability threshold in **Figure 3 (e)** and the spinwave dispersions in **Figure 4 (b)**. See Supplementary Materials for additional details.